\documentclass{iaus}
\usepackage{graphicx}
\usepackage{color}

\title[Emission Line Survey of Large PNe] 
{An Optical Emission Line Survey of Large Planetary Nebulae}

\author[Madsen et al.]   
{G.J.~Madsen$^1$, D.J.~Frew$^2$, Q.A.~Parker$^{1,2}$, \break R.J.~Reynolds$^3$ \and L.M.~Haffner$^3$}

\affiliation{$^1$NSF MPS-DRF Fellow, Anglo-Australian Observatory,  Epping, NSW 2121, Australia
\break email: madsen@aao.gov.au\\[\affilskip]
$^2$Department of Physics, Macquarie University, Sydney, NSW 2109, Australia \\[\affilskip]
$^3$Department of Astronomy, University of Wisconsin,  Madison, WI 53706, USA
}

\pubyear{2006}
\volume{234}  
\jname{Planetary Nebulae in Our Galaxy and Beyond}
\editors{M. J. Barlow \& R. H. M\'endez, eds.}

\newcommand{\ha}{H$\alpha$}
\newcommand{\hii}{{\sc{H II}}}
\newcommand{\nii}{{[\sc{N II}]}}
\newcommand{\oiii}{{[\sc{O III}]}}
\newcommand{\kms}{km~s$^{-1}$}

\begin{document}

\maketitle

\begin{abstract}
Accurate emission line fluxes from planetary nebulae (PNe) provide important constraints on the nature of the final phases of stellar evolution.  Large, evolved PNe may trace the latest stages of PN evolution, where material from the AGB wind is returned to the interstellar medium. However, the low surface brightness and spatially extended emission of large PNe have made accurate measurements of line fluxes difficult with traditional long-slit spectroscopic techniques. Furthermore, distinguishing these nebulae from \hii\ regions, supernova remnants, or interstellar gas ionized by a hot, evolved stellar core can be challenging. Here, we report on an ongoing survey of large Galactic PNe ($r > 5^\prime$) with the Wisconsin H-Alpha Mapper (WHAM), a Fabry-Perot spectrograph designed to detect faint diffuse optical emission lines with high sensitivity and spectral resolution.  Our sample includes newly revealed \ha\ enhancements from the AAO/UKST and WHAM \ha\ surveys of Parker et al.~and Haffner et al.  We present accurate emission line fluxes of \ha, \nii$\lambda$6583, and \oiii$\lambda$5007, and compare our data to other measurements.   We use the emission line ratios and kinematics of the ionized gas to assess, or in some cases reassess, the identification of some nebulae.
\keywords{Planetary nebula: general, ISM: general, HII regions, surveys}
\end{abstract}

\small

\vspace{-.6cm}
\section{The Survey}

The targets chosen for this study were observed in the lines of \ha, \nii, and \oiii\ and comprise objects cataloged as PNe in the northern sky with angular diameters generally greater than 5$^{\prime}$.  This list was supplemented with newly discovered \ha-bright nebulae from the AAO/UKST, SHASSA, and WHAM \ha\ sky surveys whose identification is uncertain (Parker et al. 2005, Haffner et al. 2003, Reynolds et al. 2005, Frew et al., this volume).  
Our new observations were carried out with the Wisconsin H-Alpha Mapper (WHAM), a remotely controlled dual-etalon Fabry-Perot facility located at Kitt Peak National Observatory.  WHAM measures an average spectrum over a 1$^{\circ}$ circular field of view, with 12 \kms\ spectral resolution over a 200 \kms\ spectral range. The observations employed an ON-OFF technique, in which two directions close to the target were also observed. The OFF spectra were subtracted from the ON spectrum to isolate emission associated with the target and remove the potential contamination from atmospheric and interstellar emission.  To calibrate the data, several bright PNe with well-known integrated fluxes and radial velocities were also observed. 

\begin{table}
\begin{minipage}{6.5cm}
\includegraphics[scale=0.6]{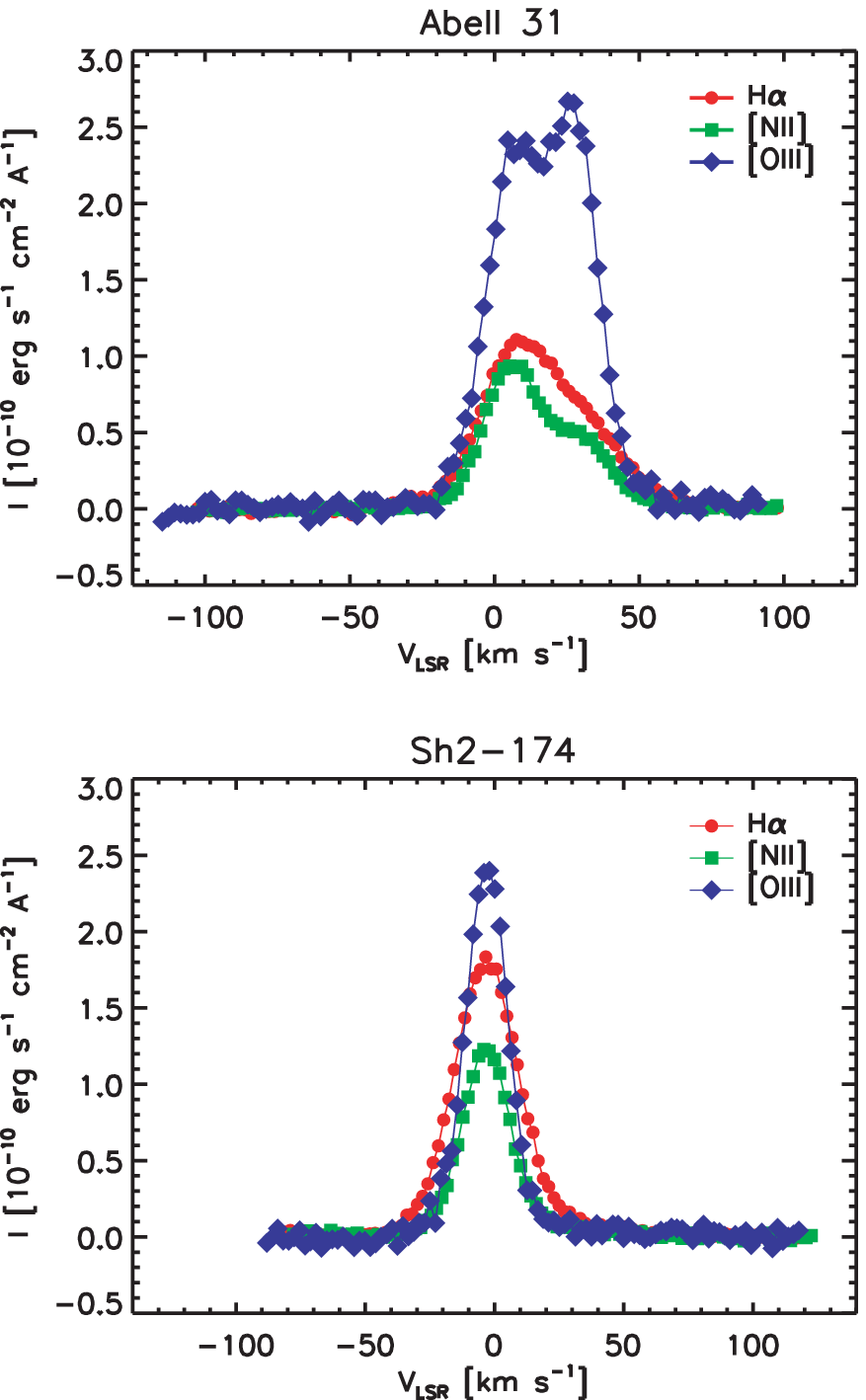}

{{\bf{Figure 1.}} Selected spectra of two targets \\in the survey. Abell 31 is a bona fide PN, \\ Sh 2-174 is not (see text).}
\end{minipage}
\begin{minipage}{6.5cm}
\caption{Selected Survey Data}
\begin{tabular}{lrrrrrrrl}
\hline
Name  & V & 
F$_{H\alpha}$ & F$_{[NII]}$ & F$_{[OIII]}$ & PN?  \\ 
\hline
Abell 24  &  1  &  -10.68  &  -10.01  &  -10.88 & Yes \\ 
Abell 28  &  0  &  -10.77  &  -11.55  &  -11.62 & Yes \\ 
Abell 31  &    &  -9.95  &  -10.10  &  -9.77 & Yes  \\ 
DHW 5  &  -5   &  -9.98  &  -10.30  &  -10.47 & No \\ 
EGB 6  &  -7  &  -10.78  &  -11.24  &  -10.78  &  Yes \\ 
FP 1824  &  1  &  -10.37   &  -10.51  &  -10.78 & Yes  \\ 
Fr 0456  &  6  &  -10.70  &  -10.66  &  -10.64  & ? \\ 
Fr 1423  &  -18 &  -10.22  &  -10.42  &  -10.04 & ? \\ 
Fr 1509  &  10  &  -10.72  &  -10.75  &  -11.16 & ?  \\ 
Fr 2027  &  17  &  -10.28  &  -10.43  &  -10.59 & ?  \\ 
Fr 2311  &  -5  &  -10.64  &  -10.70  &  -11.04 & ?  \\ 
HDW 3  &  -24  &  -10.62  &  -10.67  &  -10.62 & Yes  \\ 
Hewett 1  &  -4  &  -10.14  &  -10.50  &  -9.57 & No  \\  
IsWe 1  &  -13  &  -10.51  &  -10.81  &  -11.02 & Yes  \\ 
Jones 1  &  -10  &  -10.66  &  -11.28  &  -10.31 & Yes  \\ 
PG 0108  &  -10  &  -10.64  &  -10.83  &  -11.00 & No  \\ 
PG 0109  &  -4  &  -11.08  &  -11.02  &  -11.53 & No  \\ 
PHL 932  &  -5   &  -10.67  &  -11.32  &  & No   \\ 
Sh 2-174  &  -3  &  -9.87  &  -10.14  &  -10.05 & No  \\ 
Sh 2-188  &  -21 &  -10.38  &  -10.44  &  -10.51 & Yes \\ 
Sh 2-216  &  8  &  -9.12  &  -9.14  &  -9.19 & Yes \\ 
WPS 46  &  -52  &  -10.47  &  -10.63  &  -10.60 & ?  \\ 
WPS 60  &  -2  &  -10.22  &  -10.46  &  -10.97 & ?  \\ 
WPS 75  &     &  -10.69  &  -11.11  &  -12.09  & ? \\ 
\hline
\end{tabular}
\end{minipage}
\end{table}

	Figure 1 shows a sample of two spectra from the survey.  The line profile shapes and strength demonstrate the spectral resolution and sensitivity of the observations.  A partial summary of our results to date is shown in Table 1. This table includes the centroid radial velocity in units of \kms\ (LSR), followed by the \ha, \nii, and \oiii\ logarithmic flux in units of erg~cm$^{-2}$~s$^{-1}$, if measured.  The uncertainty in the data is less than 0.1 dex in flux and 5 \kms\ in radial velocity.  Our data have the advantage of being reduced and observed in a uniform manner, and providing absolute flux measurements and kinematics of multiple emission lines from a large sample of a unique class of PNe.  These large, evolved systems are critical in understanding PN evolution, as well as evaluating the PN luminosity function and surface brightness-radius relationship (Frew \& Parker, this volume).

\section{Assessing the Identity of Large PNe}

The large angular size and low surface brightness of many nebulae can pose a challenge in identifying them observationally as {\it{bona fide}} PNe.  Some of them may be \hii\ regions or interstellar material ionized by hot, evolved stellar cores. We use several benchmarks, including the identification of a central star and its evolutionary status, the morphology of the nebula in different emission lines, the kinematics of the central star and ionized gas, and the properties of the emission lines to distinguish PN from non-PN (see Frew et al., this volume).  We have identified 7 nebulae in our survey to date, some shown in Table 1, whose identities as PNe are suspect based on some of these criteria.  The nebular emission from Hewett 1, for example, which is claimed to be the ``largest known planetary nebula in the sky'' (Hewett et al. 2003) is probably interstellar in origin.  Chu et al. (2004) explored the puzzling emission of similar size in the vicinity of a hot white dwarf KPD 005+5106 and concluded that is likely to be interstellar as well.  Our new data will be used in conjunction with other lines of evidence in making similar assessments of other suspect PNe.  

Some of the PNe in Table 1 are newly discovered \ha-bright nebulae whose nature is uncertain; many of them lack cataloged hot stars in their vicinity.  As the survey progresses to include more of these kinds of objects, including other emission lines, we hope to improve our understanding of these objects and a more complete population of evolved PNe in the Galaxy.

This research has been generously supported by the National Science Foundation through grants AST-0204973 and AST-0401416.

\end{document}